\documentclass[letter,fleqn,usenatbib]{mnras}

\usepackage{newtxtext,newtxmath}
\usepackage[T1]{fontenc}
\usepackage{ae,aecompl}

\usepackage{graphicx}	% Including figure files
\usepackage{amsmath}	% Advanced maths commands
\usepackage{amssymb}	% Extra maths symbols
\usepackage{CJK}
\newcommand{\bdv}[1]{\mbox{\boldmath$#1$}}
\newcommand{\e}{{\rm E}}
\newcommand{\rel}{{\rm rel}}
\newcommand{\au}{{\rm au}}
\newcommand{\lens}{{\rm L}}
\newcommand{\source}{{\rm S}}
\newcommand{\mas}{{\rm mas}}
\newcommand{\masyr}{{\rm mas\,yr^{-1}}}

\begin{document}
\begin{CJK*}{UTF8}{gbsn}

%%%%%%%%%%%%%%%%%%% TITLE PAGE %%%%%%%%%%%%%%%%%%%

% Title of the paper, and the short title which is used in the headers.
% Keep the title short and informative.
\title[Isolated black holes from microlensing]{Detecting isolated stellar-mass black holes in the absence of microlensing parallax effect}

% The list of authors, and the short list which is used in the headers.
% If you need two or more lines of authors, add an extra line using \newauthor
\author[Karolinski \& Zhu]{
Numa Karolinski$^{1, 2}$\thanks{E-mail: numa.karolinski@mail.mcgill.ca}
and Wei Zhu (祝伟)$^{1}$\thanks{E-mail: weizhu@cita.utoronto.ca}
\\
% List of institutions
$^{1}$Canadian Institute for Theoretical Astrophysics, University of Toronto, 60 St. George Street, Toronto, ON M5S 3H8, Canada\\
$^{2}$Department of Physics, McGill University, 845 Sherbrooke St W, Montreal, Quebec H3A 0G4, Canada
}

% These dates will be filled out by the publisher
\date{Accepted XXX. Received YYY; in original form ZZZ}

% Enter the current year, for the copyright statements etc.
\pubyear{2020}

% Don't change these lines
\label{firstpage}
\pagerange{\pageref{firstpage}--\pageref{lastpage}}
\maketitle

\begin{abstract}
Gravitational microlensing can detect isolated stellar-mass black holes (BHs), which are believed to be the dominant form of Galactic BHs according to population synthesis models. Previous searches for BH events in microlensing data focused on long-timescale events with significant microlensing parallax detections. Here we show that, although BH events preferentially have long timescales, the microlensing parallax amplitudes are so small that in most cases the parallax signals cannot be detected statistically significantly. We then identify OGLE-2006-BLG-044 to be a candidate BH event because of its long timescale and small microlensing parallax. Our findings have implications to future BH searches in microlensing data.
\end{abstract}

\begin{keywords}
gravitational lensing: micro -- stars: black holes -- methods: data analysis
\end{keywords}

%%%%%%%%%%%%%%%%%%%%%%%%%%%%%%%%%%%%%%%%%%%%%%%%%%

%%%%%%%%%%%%%%%%% BODY OF PAPER %%%%%%%%%%%%%%%%%%
\section{Introduction}

Stars with initial masses $\gtrsim20\,M_\odot$ are believed to end their lives in stellar-mass BHs. It has been estimated that the Milky Way contains $\sim10^8$ stellar-mass BHs \citep[e.g.,][]{Shapiro:1983}. Although most massive stars are found in binaries or higher multiples, stellar evolutions frequently lead to mergers (due to common envelope evolution or gravitational radiation) or disruptions of stellar binaries, resulting in the majority of stellar-mass BHs in isolation \citep[e.g.,][]{Belczynski:2002, Belczynski:2004, Olejak:2019, Wiktorowicz:2019}. Therefore, isolated BHs are important for our understanding of the stellar-mass BH population. In particular, a statistical knowledge of isolated BHs can help to constrain the formation channels of gravitational wave sources, such as those found by LIGO and Virgo \citep[e.g.,][]{Abbott:2016, Abbott:2019}.

While many techniques exist to detect BHs in binaries, gravitational microlensing is perhaps the only viable technique to detect isolated BHs \citep{Einstein:1936, Paczynski:1986}. Starting from the widely used initial mass function and the general criterion for BH formation, \citet{Gould:2000} estimated that $\sim1\%$ of microlensing events towards the bulge direction should be due to stellar-mass BHs. Later studies that implemented more complicated physics and/or focused on specific surveys or observing strategies found consistent results \citep[e.g.,][]{Oslowski:2008, Rybicki:2018, Wiktorowicz:2019, Lam:2020}. \citet{Wiktorowicz:2019} used the stellar population synthesis code, \texttt{StarTrack}, and estimated that there should be 14 and 26 BH microlensing events per year in surveys like OGLE-III and OGLE-IV, respectively. \citet{Lam:2020} developed a new population synthesis code specifically for the searches of compact object microlensing events, \texttt{PopSyCLE}. The authors found that with current ground-based astrometric follow-ups \citep[e.g.,][]{Lu:2016} of events from existing microlensing surveys one should expect to detect one to two BHs each year.

One issue with using microlensing to detect BHs is how to identify them from the much more abundant normal microlenses, namely, stars. In the standard microlensing model \citep{Paczynski:1986} only the event timescale, given by
\begin{equation}
t_\e \equiv \frac{\theta_\e}{\mu_\rel} = \frac{\sqrt{\kappa M_\lens \pi_\rel}}{\mu_\rel},
\end{equation}
is related to the lens mass $M_\lens$. Here $\theta_\e$ is the angular Einstein radius, $\mu_\rel$ and $\pi_\rel$ are the relative proper motion and relative parallax between the lens and source, respectively, and $\kappa$ is a constant
\begin{equation}
\kappa \equiv \frac{4G}{c^2 \au} \approx 8.14 \frac{\rm mas}{M_\odot} .
\end{equation}
With one observable ($t_\e$) and three unknowns ($M_\lens$, $\mu_\rel$, and $\pi_\rel$), one cannot uniquely determine the lens mass. Statistical studies of the event timescale distribution have been done to infer the mass function from substellar objects up to stellar-mass BHs \citep[e.g.,][]{Sumi:2011, Mroz:2017}, but results from this approach are very much subject to the details of the adopted galactic model.

\begin{figure}
\centering
\includegraphics[width=0.95\columnwidth]{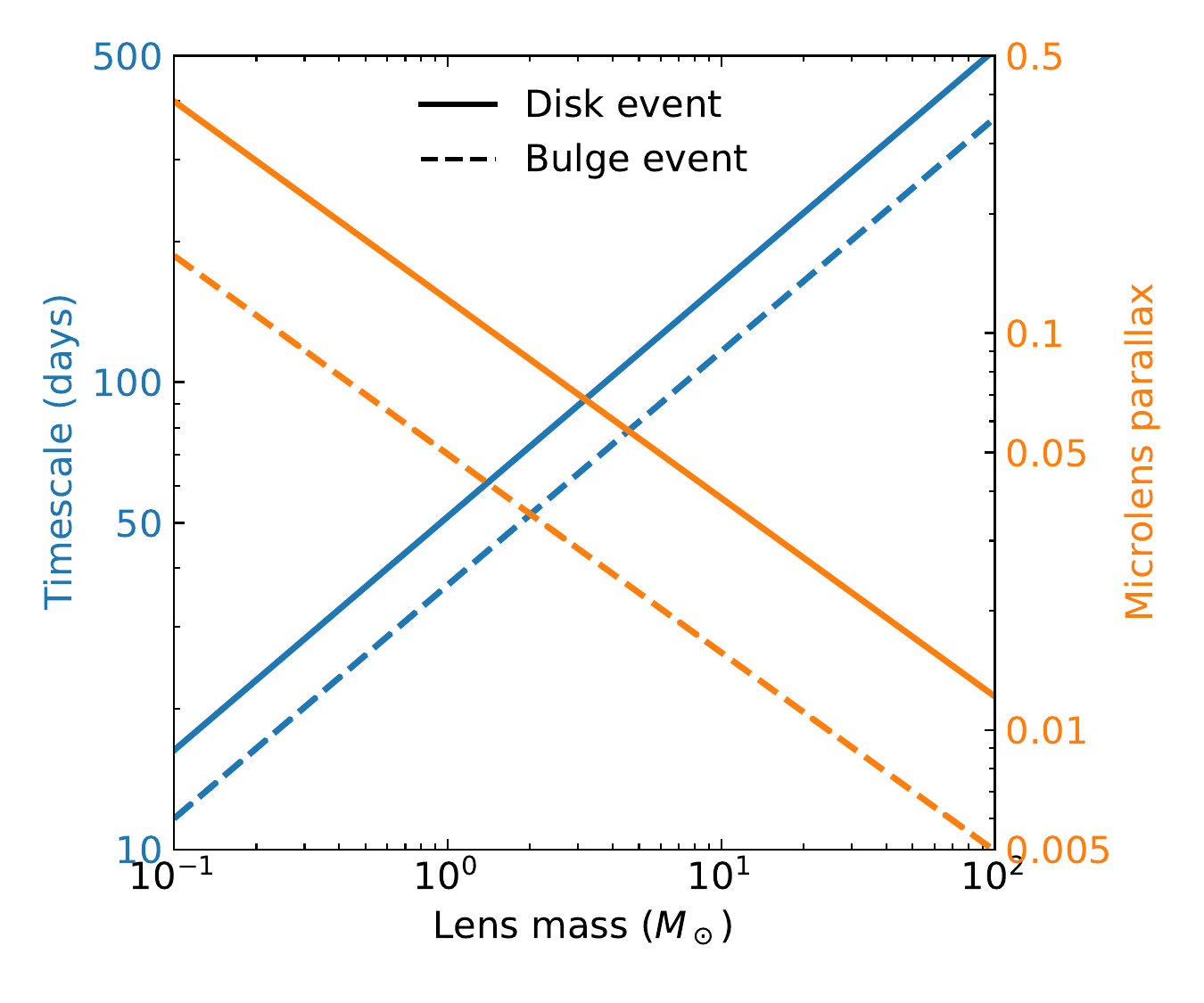}
\caption{The event timescale $t_\e$ (blue) and the microlensing parallax parameter $\pi_\e$ (orange) as functions of lens mass for typical disc events ($\pi_\rel=0.12\,\mas$, $\mu_\rel=7\,\masyr$) and typical bulge events ($\pi_\rel=0.02\,\mas$, $\mu_\rel=4\,\masyr$), respectively. Microlensing events arisen from isolated stellar-mass BHs ($\gtrsim3~M_\odot$) should have large $t_\e$ but small $\pi_\e$.
\label{fig:dependence}}
\end{figure}

Once measured (or constrained), the microlensing parallax parameter, given by \citep{Gould:2000b}
\begin{equation}
\pi_\e \equiv \frac{\pi_\rel}{\theta_\e} ,
\end{equation}
can reduce the degree of freedom in the problem. As $t_\e \propto M_\lens^{1/2}$ whereas $\pi_\e \propto M_\lens^{-1/2}$ (see Figure~\ref{fig:dependence}), the combination of the two can significantly reduce the statistical uncertainty in inferred lens mass, even though the mass still cannot be uniquely determined \citep{Han:1995, Zhu:2017}. 

With only ground-based observations, $\pi_\e$ is constrained through the annual parallax effect that originates from the orbital motion of the Earth around the Sun \citep{Gould:1992}. Such an effect is only detectable in relatively long-timescale ($t_\e \gtrsim {\rm yr}/2\pi$) events. Luckily, events due to stellar-mass BHs belong to such a category. Since the early days of microlensing searches, efforts have been taken to identify events with long timescales and annual parallax effect and to infer masses of the foreground lenses \citep[e.g.,][]{Mao:2002, Bennett:2002, Agol:2002, Poindexter:2005, Wyrzykowski:2016}. Here we use \citet{Wyrzykowski:2016} as a representative example to demonstrate the standard procedure. Starting from over 3600 microlensing events found by the OGLE-III survey \citep{Udalski:2008, Wyrzykowski:2015}, \citet{Wyrzykowski:2016} found 59 parallax events. These were the events that showed significant ($>50$) $\chi^2$ improvement in the light curve fit after parallax parameters were included. From the standard Bayesian statistical inference the authors identified 13 microlenses consistent with being stellar remnants.

For such massive microlenses as the isolated stellar-mass BHs, the long timescale makes it easier to detect the annual parallax effect, but the reduced amplitude of the parallax parameter also makes the parallax effect more subtle (see Figure~\ref{fig:dependence}). We show in Section~\ref{sec:detectability} that the parallax effect in a truly BH event is mostly undetectable. As such, we re-analyze the OGLE-III dataset with a method slightly different from that of \citet{Wyrzykowski:2016} and identify one candidate BH event. This search is presented in Section~\ref{sec:search}. We discuss the results in Section~\ref{sec:discussion}.

%%%%%%%%%%
\section{Annual parallax effect in a BH event is undetectable} \label{sec:detectability}

With a single lens, the magnified source flux at any given time $t$ is given by
\begin{equation}
F(t) = F_\source [A(t) -1] + F_{\rm base} ,
\end{equation}
where the magnification $A$ is given by
\begin{equation}
A(t) = \frac{u^2+2}{u\sqrt{u^2+4}} .
\end{equation}
Here $F_\source$ is the source flux at baseline ($A=1$), $F_{\rm base}$ is the total flux (i.e., source and blend) at baseline, and $u$ is the dimensionless separation between the source and the lens. When annual parallax is included \citep{Gould:2004}
\begin{equation}
u^2 = \left(\frac{t-t_0}{t_\e} + \pi_{\rm E,N} s_{\rm N} + \pi_{\rm E,E} s_\e \right)^2 + (u_0 + \pi_{\rm E,N} s_\e - \pi_{\rm E,E} s_{\rm N})^2 ,
\end{equation}
where $t_0$ is the event peak time (in the absence of parallax effect), $u_0$ is the impact parameter, $t_\e$ is the event timescale in the geocentric reference frame, $\pi_{\rm E,N}$ and $\pi_{\rm E,E}$ are the parallax components along the north and east direction, respectively. The quantities $s_{\rm N}$ and $s_{\rm E}$ are the corresponding offsets (in units of au) between Earth's actual position and the position of Earth in the absence of parallax effect. These offsets are evaluated at a fixed time very close to the peak time $t_0$ and are thus independent of the details of the event.

The detectability of the parallax effect, namely the deviations of $\pi_{\rm E,N}$ and $\pi_{\rm E,E}$ from zeros, can be evaluated through a Fisher matrix analysis. With the nuisance parameter $F_{\rm base}$ ignored, the microlensing light curve is described by the parameter set $\bdv{\theta}\equiv(t_0,\,u_0,\,t_\e,\,\pi_{\rm E,N},\,\pi_{\rm E,E},\,F_\source)$. The Fisher information matrix is then
\begin{equation}
\mathcal{F}_{ij} = \sum_{\{t_k\}} \frac{1}{\sigma_F^2(t_k)} \frac{\partial F(t_k)}{\partial \theta_i} \frac{\partial F(t_k)}{\partial \theta_j} .
\end{equation}
Here $\sigma_F$ denotes the uncertainty in measured flux. The summation is done over a time series $\{t_k\}$. The inverse of the Fisher information matrix gives the covariance matrix of $\bdv{\theta}$. The detectability of the parallax effect can be quantified as
\begin{equation} \label{eqn:chi2}
\Delta \chi^2 = (\bdv{\theta}_0 -\bdv{\theta}) \mathcal{F} (\bdv{\theta}_0 - \bdv{\theta})^{\rm T} ,
\end{equation}
where the no-parallax parameter set $\bdv{\theta}_0$ has $(\pi_{\rm E,N},\,\pi_{\rm E,E})=(0,\,0)$ and other parameters the same as those in $\bdv{\theta}$.

\begin{figure*}
\includegraphics[width=0.9\textwidth]{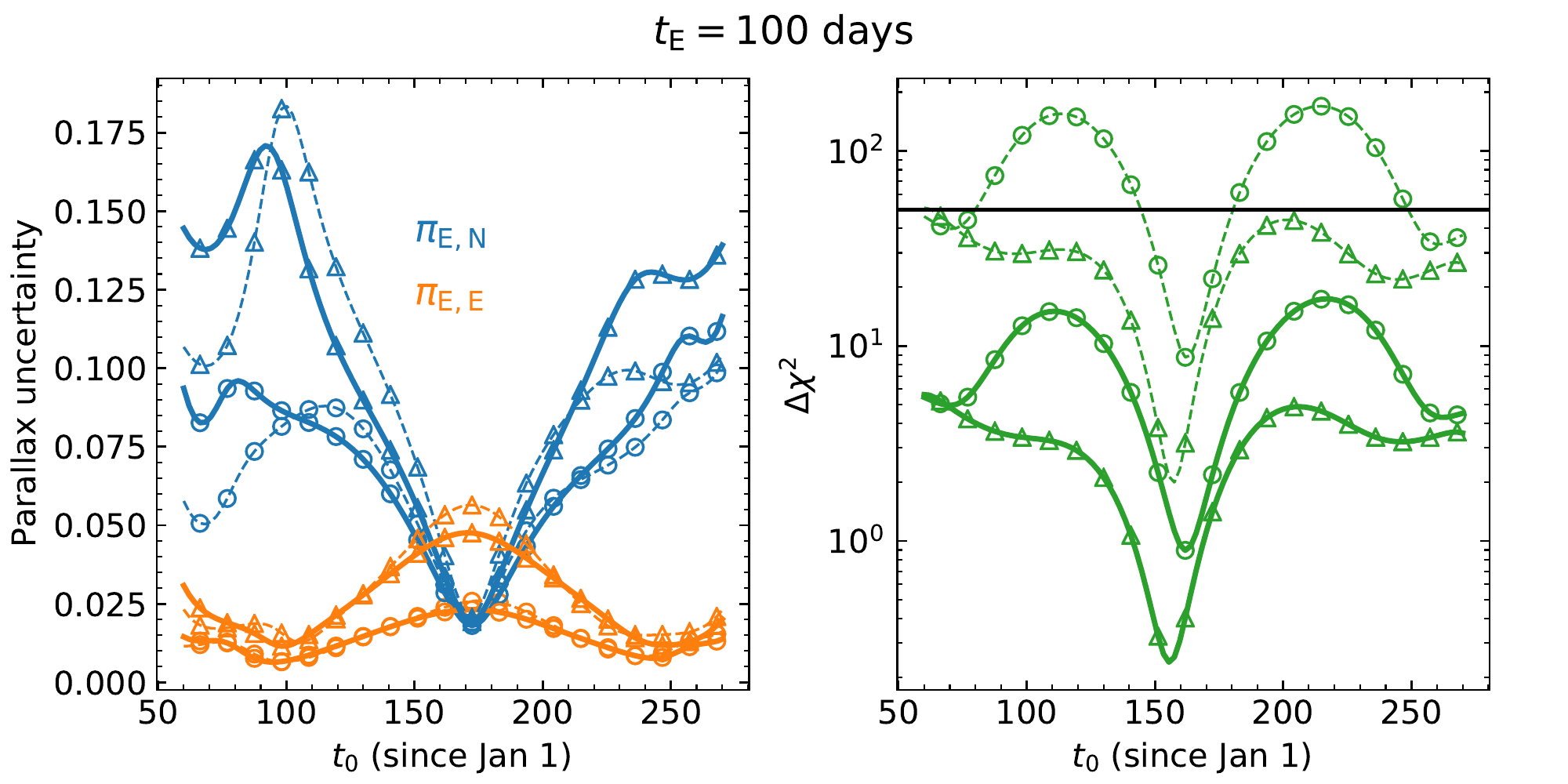}
\caption{\textit{Left panel:} Uncertainties on the two components of the microlensing parallax vector, $\pi_{\rm E,N}$ (blue) and $\pi_{\rm E,E}$ (orange), from different event setups. The $x-$axis denotes the event peak time $t_0$. The parallax amplitude $\pi_\e$ is $0.1$ for the dashed curves and $0.03$ for the solid curves. The latter is more typical for BH events. The impact parameter $u_0$ is $0.1$ for the circles and $0.3$ for the triangles. The event timescale $t_\e$ is chosen to be 100 days and the source baseline magnitude $I=18$ mag. \textit{Right panel:} The resulting detectability of the parallax effect, measured in $\Delta \chi^2$ (Equation~\ref{eqn:chi2}), for different event setups. The black horizontal line marks $\Delta\chi^2=50$, the threshold used in \citet{Wyrzykowski:2016} for selecting BH-candidate events.
\label{fig:detectability}}
\end{figure*}

We compute $\Delta \chi^2$ values for typical isolated BH events in an OGLE-like microlensing survey. A typical event timescale is chosen, $t_\e=100$ days (e.g., Figure~13 of \citealt{Lam:2020}), and two sets of the parallax vector are explored, $(\pi_{\rm E, N},\,\pi_{\rm E,E})=(0.1/\sqrt{2},\,0.1/\sqrt{2})$ and $(0.03/\sqrt{2},\,0.03/\sqrt{2})$. While the former is chosen for comparison purposes, the latter is more typical for BH events (see Figure~\ref{fig:dependence}). We also fix the source magnitude to $I=18$ and assume no blending, which are not atypical for the OGLE-III events \citep{Wyrzykowski:2015}. To mimic the OGLE-III survey, we assume that the the flux noise is sky-limited (i.e., $\sigma_F=$constant and thus $\sigma_I\propto F^{-1}$) and the magnitude error $\sigma_I=0.04$ at $I=18$ mag source.
\footnote{Such a simple assumption fails for very bright measurements ($I\lesssim16$). However, for our chosen event parameters its impact is very limited.}
The cadence is chosen to be 1/day. With two representative values of the impact parameter, $u_0=0.1$ and $0.3$, and different values of the event peak time, we can then compute the uncertainties on the parallax components and the detectability of the parallax effect. The results are shown in Figure~\ref{fig:detectability}.

As shown in Figure~\ref{fig:detectability}, both components of the parallax vector show large variations with the event peak time $t_0$, with the eastern component, $\pi_{\rm E,E}$, usually better constrained than the northern component, $\pi_{\rm E,N}$. The event peak time $t_0$ determines the orientation of the sky-projected acceleration of the Earth from the Sun relative to the Earth's trajectory. The eastern component preferentially introduces asymmetry into the microlensing light curve whereas the northern component preferentially changes the overall magnification \citep[e.g.,][]{Smith:2003}. The two effects switch when the projected acceleration of the Earth from the Sun is perpendicular to Earth's trajectory (i.e., at conjunction point). Additionally, the relatively small parallax amplitudes we have chosen have a rather minor effect on the parallax uncertainties, whereas the different impact parameters, through the larger impact on the magnifications, can lead to considerate differences in the resulting parallax uncertainties.

The main point of this exercise is to show that the microlensing parallax effect in a typical BH event (with $t_\e=100$ days and $\pi_\e=0.03$) is usually undetectable according to the criteria of \citet{Wyrzykowski:2016} ($\Delta\chi^2>50$), regardless of the exact choices of other values.

%%%%%%%%%%%%%%%%%%%%%%%%%%%%%%%%
\section{A BH candidate event with no parallax detection} \label{sec:search}

\begin{figure}
\includegraphics[width=0.9\columnwidth]{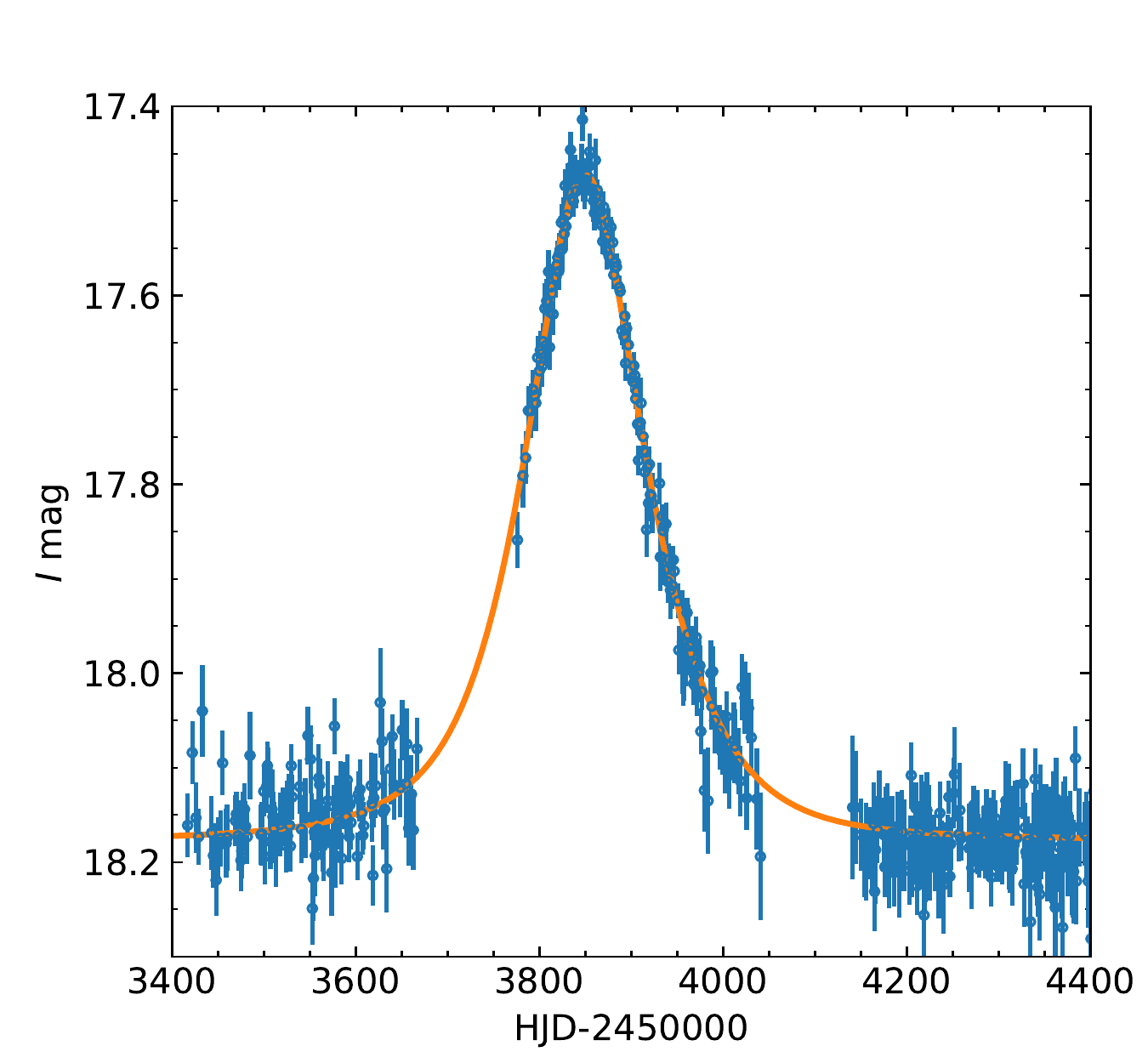}
\caption{Light curve of the microlensing event OGLE-2006-BLG-044. The best-fit model is shown as orange curve.
\label{fig:lc}}
\end{figure}

\begin{figure*}
\includegraphics[width=0.95\textwidth]{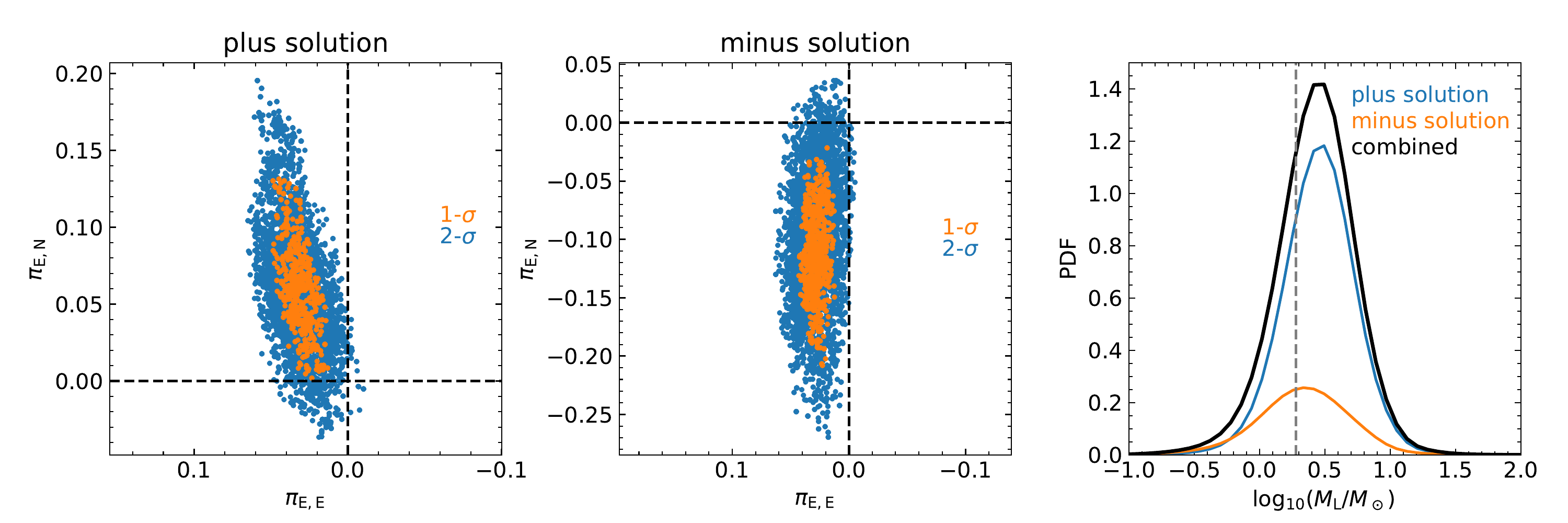}
\caption{Posterior distributions of the parallax vector for the plus (left) and minus (middle) solutions. The dashed lines indicate the zero parallax positions. The probability distributions of the lens mass are shown in the right panel. Contributions from individual solutions are indicated. The grey dashed vertical line indicates $M_{\rm L}=1.9\,M_\odot$, the 95\% upper limit on the lens mass if the lens is luminous.
\label{fig:posterior}}
\end{figure*}

Building on the conclusion from the previous section, we revisit the OGLE-III microlensing dataset presented by \citet{Wyrzykowski:2015} and identify one event in which the lens is possibly an isolated stellar-mass BH. The identified event did not have a statistically significant parallax signal and thus was not included in the list of candidate events in \citet{Wyrzykowski:2016}.

The candidate event was found through a systematic search in the OGLE-III microlensing sample. With the microlensing parameters reported by \citet{Wyrzykowski:2015}, we first selected events with relatively long timescales and reasonably well sampled light curves. These events are then fitted for microlensing parallax effect using the Markov Chain Monte Carlo (MCMC) method. The parallax model is generated by the \texttt{MulensModel} package \citep{Poleski:2019} and the MCMC is done with the \texttt{emcee} package \citep{ForemanMackey:2013}. The resulting Markov chains are visually inspected for convergence and the converged chains are saved for the mass inference. Two degenerate solutions due to the ecliptic degeneracy \citep{Skowron:2011} are identified. They are labeled ``plus'' (for $u_0>0$) and ``minus'' (for $u_0<0$) solutions, respectively.

Our method to infer the lens mass is similar to that of \citet{Zhu:2017} but with some modifications. In a short summary, the method derives the probability distribution of the lens mass in a Bayesian framework \citep[e.g.,][]{Batista:2011}
\begin{equation} \label{eqn:bayes}
P(\log M_{\rm L}) \propto \int \frac{d^4 \Gamma}{d t_{\rm E} d \log{M_{\rm L}} d^2 \bdv{\tilde{v}_{\rm hel}}} \mathcal{L}(t_{\rm E},\,\bdv{\tilde{v}_{\rm hel}}) dt_{\rm E} d^2 \bdv{\tilde{v}_{\rm hel}} ,
\end{equation}
where
\begin{equation}
    \frac{d^4 \Gamma}{d t_{\rm E} d \log{M_{\rm L}} d^2 \bdv{\tilde{v}_{\rm hel}}} = 4 n_{\rm L,\star} D_{\rm L}^4 f_v(\bdv{\tilde{v}_{\rm hel}}) \frac{d \xi\left(M_{\mathrm{L}}\right)}{d \log M_{\mathrm{L}}} \frac{\mu_{\rm rel}^3}{\tilde{v}_{\rm hel}}
\end{equation}
represents the prior information combining the lens mass and lens kinematics. Here $n_{\rm L,\star}$ is the number density of the lens at a given location, $f_v(\bdv{\tilde{v}_{\rm hel}})$ is the probability distribution of the transverse velocity vector $\bdv{\tilde{v}_{\rm hel}}$, $d\xi/d\log{M_{\rm L}}$ is the lens mass function, and $\mathcal{L}(t_\e^\prime,\,\bdv{\tilde{v}_{\rm hel}})$ is the likelihood distribution of the microlensing parameters from the light curve modeling. Note that we use $t_\e^\prime$ for the event timescale measured in the heliocentric frame, which can be determined for a given parameter set $\bdv{\theta}$. The integral in Equation~(\ref{eqn:bayes}) is done with a summation over the entire Markov chain.

For nearly all of the events modeled here, the amplitude of the parallax vector is statistically consistent with zero and thus the direction, which is also the direction of $\bdv{\tilde{v}_{\rm hel}}$, is nearly unconstrained. To prevent the prior mass function from driving the posterior to very unusual lens kinematics, we use a log-flat distribution as the lens mass function, $d\xi/d\log{M_{\rm L}} \propto$ constant. Such a mass function can produce unbiased lens mass probability distribution, as is demonstrated in Appendix~\ref{sec:prior}. The upper limit of the mass function is extended up to $100~M_\odot$. Other parts of the Galactic model are detailed in \citet{Zhu:2017}.

\begin{table}
\caption{Fitting parameters of the microlensing event OGLE-2006-BLG-044. Here $I_{\rm b}$ is the $I$-band magnitude of the blend object. These parameters are measured in the geocentric reference frame and the parallax reference time is set at HJD$=2,453,849$. The reduced $\chi^2$ (i.e., $\chi^2$ per degree of freedom) values are also listed. For a reference, the standard model yields $\chi^2=924.8$ for 890 degrees of freedom.
\label{tab:parameters}}
\centering
\begin{tabular}{ccc}
\hline
 & Plus solution & Minus solution \\
\hline
$\chi^2$/d.o.f. & 921.7/888 & 920.8/888 \\
$t_0-2450000$ & $3848.8\pm0.5$ & $3848.8\pm0.5$ \\
$u_0$ & $0.54\pm0.09$ & $-0.58\pm0.06$ \\
$t_{\rm E}$ (days) & $106\pm12$ & $99\pm6$ \\
$\pi_{\rm E,N}$ & $0.06\pm0.05$ & $-0.10\pm0.07$ \\
$\pi_{\rm E,E}$ & $0.034\pm0.018$ & $0.028\pm0.017$ \\
$I_{\rm b}$ $^{\rm a}$ & $>18.9$ & $>19.5$ \\
\hline
\end{tabular} \\
$^{\rm a}$ Derived from the 95\% upper limit on the blend flux.
\end{table}

The best candidate we identified is OGLE-2006-BLG-044
\footnote{This event is also labeled as OGLEIII-ULENS-1643\ in the \citet{Wyrzykowski:2015} catalog.}.
The light curve is shown in Figure~\ref{fig:lc} and our best-fit parameters are given in Table~\ref{tab:parameters}. As seen from Table~\ref{tab:parameters} as well as illustrated in the left and middle panels of Figure~\ref{fig:posterior}, the parallax parameters of this event are consistent with zero at the 2-$\sigma$ level. If the lens is luminous, it would contribute to the blend flux. In other words, the blend flux provides an upper limit on the mass of a luminous lens. With the distance modulus (14.58) and the extinction ($A_I=2.56$) given by the OGLE survey at this particular line of sight \citep{Nataf:2013}, we can convert the apparent $I$-band magnitude of the blend object into the absolute magnitude $M_I$. With the stellar mass-absolute magnitude relations from \citet{Pecaut:2012} and \citet{Pecaut:2013} we then find that the mass of the luminous lens cannot be higher than $1.6\,M_\odot$ or $1.9\,M_\odot$ for the minus and plus solutions, respectively. Both are the 95\% upper limit. This calculation has assumed that the lens is behind all the dust and very close to the source, which is assumed to be in the bulge. The mass limit for a luminous lens can only be lower if the lens gets closer.

As seen from Figure~\ref{fig:posterior}, although the parallax is not statistically significantly detected in OGLE-2006-BLG-044, the inferred lens mass distribution suggests that the lens is relatively massive and probably (with 66\% probability) dark. If $3\,M_\odot$ is taken as the upper limit of any neutron star, then there is a 39\% probability that the lens in OGLE-2006-BLG-044 is a stellar-mass BH.

A similar Bayesian analysis is performed on the lens distance. The median value of the lens distance probability distribution is found to be 3.2 kpc and the 68\% confidence interval is 2.0-4.9 kpc. Therefore, the lens of OGLE-2006-BLG-044 is likely in the disc.

%%%%%%%%%%%%%%%%%%%%%%%%%%%%%%%%
\section{Discussion} \label{sec:discussion}

Gravitational microlensing can detect dark objects such as stellar-mass BHs. Compared to other techniques, microlensing is unique in its sensitivity to isolated BHs, which are thought to be the dominant form of stellar-mass BHs according to population synthesis models.

Previous searches for BH events in microlensing data have focused on long-timescale events with significant parallax signals. Here we show that, while BH events are indeed preferentially long-timescale, their microlensing parallax amplitudes are so small that the parallax signal becomes undetectable in the most cases.

Following this new finding, we then looked into the public OGLE-III microlensing database and identified a new BH candidate event, OGLE-2006-BLG-044. Although the parallax signal was not detected at any statistically significant level, the long timescale and the small parallax amplitude together suggest that the lens is probably dark and has a 39\% probability of being a stellar-mass BH.

Our results have implications to future searches and statistical analysis of Galactic BHs in the microlensing dataset.

\section*{Acknowledgements}

This work is the result of the 2019 Summer Undergraduate Research Program (SURP) in astronomy \& astrophysics at the University of Toronto. W.Z. was partly supported by the Beatrice and Vincent Tremaine Fellowship at CITA.

\section*{Data availability}

No new data were generated or analysed in support of this research.

%%%%%%%%%%%%%%%%%%%%%%%%%%%%%%%%%%%%%%%%%%%%%%%%%%


\appendix
\section{The choice of mass function prior} \label{sec:prior}

\begin{figure*}
\includegraphics[width=0.95\textwidth]{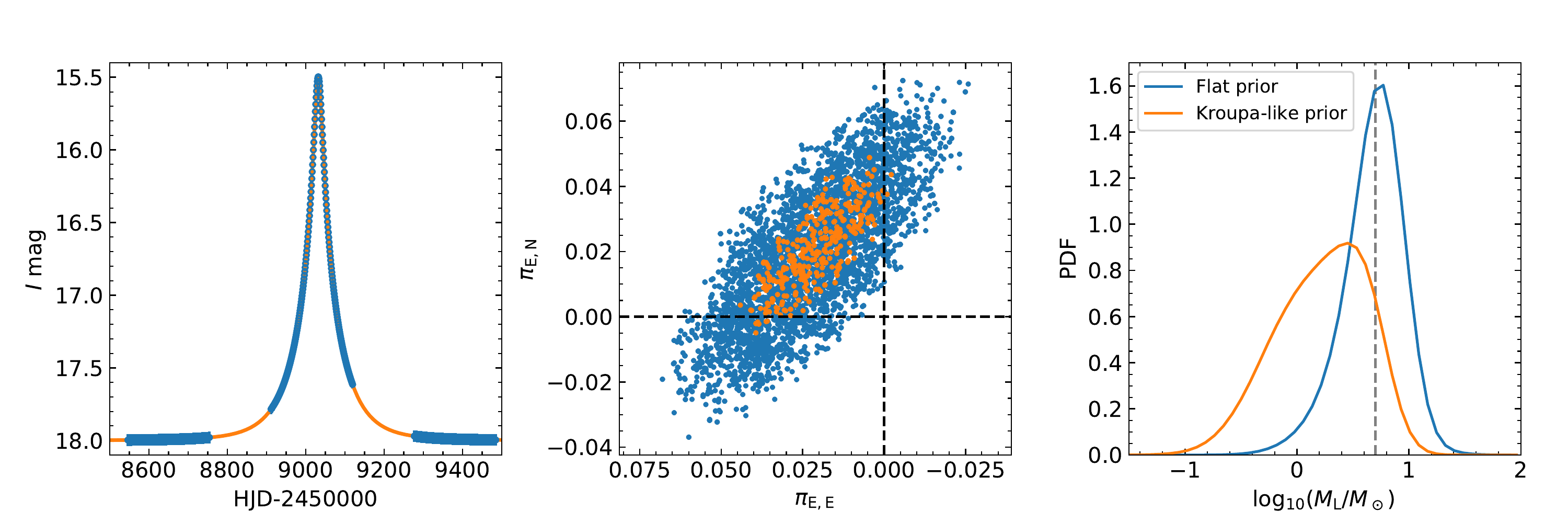}
\caption{A simulated BH event light curve (left panel), the constraint on the microlensing parallax parameters (middle panel), and the inferred lens mass probability distribution (right panel). We use $t_{\rm E}=100$ days, $u_0=0.1$, and $\pi_{\rm E,N}=\pi_{\rm E,E}=0.03/\sqrt{2}$ in the light curve simulation. Events with the chosen $t_{\rm E}$ and $\pi_{\rm E}$ are overwhelmingly dominated by BH lenses (e.g., Figure~13 of \citealt{Lam:2020}). The meanings of the data points in the middle panel are the same as those in Figure~\ref{fig:posterior}. When inferring the lens mass distribution, we have used two different mass priors. A flat prior can successfully recover the most likely mass of the lens, with the vertical dashed line marking $M_{\rm L}=5\,M_\odot$. A steep mass function that better matches the current stellar mass function, $d\xi/d\log{M_{\rm L}} \propto M_{\rm L}^{-1.7}$, which we call a Kroupa-like prior \citep{Kroupa:2003}, will lead to an underestimated lens mass probability distribution.
\label{fig:prior}}
\end{figure*}

When the lens mass is inferred in the Bayesian analysis, a flat mass function ($d\xi/d\log{M_{\rm L}}$ is constant) is used. Such a mass function is far from being representative of the current-day stellar mass function, but it provides an unbiased estimate of the lens mass in our analysis.

We use a simulated microlensing event to demonstrate this point. A event light curve was generated with $t_0=2,459,032$ (i.e., 2020 July 1st), $t_{\rm E}=100$ days, $u_0=0.1$, and $\pi_{\rm E,N}=\pi_{\rm E,E}=0.03/\sqrt{2}$, with the data sampling and precision chosen to best match that of OGLE-III (see Section~\ref{sec:detectability}). See the left panel of Figure~\ref{fig:prior} for an illustration of the simulated light curve. Galactic microlensing events with the chosen $t_{\rm E}$ and $\pi_{\rm E}$ are overwhelmingly dominated by BH events, as can be seen clearly in the Figure 13 of \citet{Lam:2020}. The simulated event was then processed in the same way as the OGLE-III events in this study to produce the constraints on the parallax parameters (middle panel of Figure~\ref{fig:prior}) and the lens mass probability distribution (right panel of Figure~\ref{fig:prior}). The inferred lens mass distribution peaks at the most probable lens mass (i.e., BH lenses). 

For a comparison, a different lens mass function was used to derive the lens mass probability distribution and the result was also shown in the right panel of Figure~\ref{fig:prior}. This Kroupa-like mass function \citep{Kroupa:2003}, with its form $d\xi/d\log{M_{\rm L}} \propto M_{\rm L}^{-1.7}$, gives more weights to the low-mass lenses, making it a better match to the current-day stellar mass function (although it is still far from perfect). However, the lens mass probability distribution inferred with the use of this more realistic mass function would suggest that the lens is most likely a normal star. The use of this Kroupa-like mass function, or any kind of realistic stellar mass function, tends to downplay the importance of the information from the parallax constraint (or measurement) and therefore leads to an underestimated lens mass probability distribution.

% Don't change these lines
\bsp	% typesetting comment
\label{lastpage}
\end{CJK*}

\begin{thebibliography}{99}
\bibitem[\protect\citeauthoryear{Abbott, et al.}{2016}]{Abbott:2016} Abbott B.~P., et al., 2016, PhRvL, 116, 061102
\bibitem[\protect\citeauthoryear{Abbott, et al.}{2019}]{Abbott:2019} Abbott B.~P., et al., 2019, PhRvX, 9, 031040
\bibitem[\protect\citeauthoryear{Agol, et al.}{2002}]{Agol:2002} Agol E., Kamionkowski M., Koopmans L.~V.~E., Blandford R.~D., 2002, ApJL, 576, L131
\bibitem[\protect\citeauthoryear{Batista, et al.}{2011}]{Batista:2011} Batista V., et al., 2011, A\&A, 529, A102
\bibitem[\protect\citeauthoryear{Belczynski, Bulik \& Klu{\'z}niak}{2002}]{Belczynski:2002} Belczynski K., Bulik T., Klu{\'z}niak W., 2002, ApJL, 567, L63
\bibitem[\protect\citeauthoryear{Belczynski, Sadowski \& Rasio}{2004}]{Belczynski:2004} Belczynski K., Sadowski A., Rasio F.~A., 2004, ApJ, 611, 1068
\bibitem[\protect\citeauthoryear{Bennett, et al.}{2002}]{Bennett:2002} Bennett D.~P., et al., 2002, ApJ, 579, 639
\bibitem[\protect\citeauthoryear{Einstein}{1936}]{Einstein:1936} Einstein A., 1936, Sci, 84, 506
\bibitem[\protect\citeauthoryear{Foreman-Mackey, et al.}{2013}]{ForemanMackey:2013} Foreman-Mackey D., Hogg D.~W., Lang D., Goodman J., 2013, PASP, 125, 306
\bibitem[\protect\citeauthoryear{Gould}{1992}]{Gould:1992} Gould A., 1992, ApJ, 392, 442
\bibitem[\protect\citeauthoryear{Gould}{2000a}]{Gould:2000} Gould A., 2000a, ApJ, 535, 928
\bibitem[\protect\citeauthoryear{Gould}{2000b}]{Gould:2000b} Gould A., 2000b, ApJ, 542, 785
\bibitem[\protect\citeauthoryear{Gould}{2004}]{Gould:2004} Gould A., 2004, ApJ, 606, 319
\bibitem[\protect\citeauthoryear{Han \& Gould}{1995}]{Han:1995} Han C., Gould A., 1995, ApJ, 447, 53
\bibitem[\protect\citeauthoryear{Kroupa \& Weidner}{2003}]{Kroupa:2003} Kroupa P., Weidner C., 2003, ApJ, 598, 1076
\bibitem[\protect\citeauthoryear{Lam, et al.}{2020}]{Lam:2020} Lam C.~Y., Lu J.~R., Hosek M.~W., Dawson W.~A., Golovich N.~R., 2020, ApJ, 889, 31
\bibitem[\protect\citeauthoryear{Lu, et al.}{2016}]{Lu:2016} Lu J.~R., Sinukoff E., Ofek E.~O., Udalski A., Kozlowski S., 2016, ApJ, 830, 41
\bibitem[\protect\citeauthoryear{Mao, et al.}{2002}]{Mao:2002} Mao S., et al., 2002, MNRAS, 329, 349
\bibitem[\protect\citeauthoryear{Mr{\'o}z, et al.}{2017}]{Mroz:2017} Mr{\'o}z P., et al., 2017, Natur, 548, 183
\bibitem[\protect\citeauthoryear{Nataf, et al.}{2013}]{Nataf:2013} Nataf D.~M., et al., 2013, ApJ, 769, 88
\bibitem[\protect\citeauthoryear{Olejak, et al.}{2019}]{Olejak:2019} Olejak A., Belczynski K., Bulik T., Sobolewska M., 2019, arXiv, arXiv:1908.08775
\bibitem[\protect\citeauthoryear{Os{\l}owski, et al.}{2008}]{Oslowski:2008} Os{\l}owski S., Moderski R., Bulik T., Belczynski K., 2008, A\&A, 478, 429
\bibitem[\protect\citeauthoryear{Paczynski}{1986}]{Paczynski:1986} Paczynski B., 1986, ApJ, 304, 1
\bibitem[\protect\citeauthoryear{Pecaut \& Mamajek}{2013}]{Pecaut:2013} Pecaut M.~J., Mamajek E.~E., 2013, ApJS, 208, 9
\bibitem[\protect\citeauthoryear{Pecaut, Mamajek \& Bubar}{2012}]{Pecaut:2012} Pecaut M.~J., Mamajek E.~E., Bubar E.~J., 2012, ApJ, 746, 154
\bibitem[\protect\citeauthoryear{Poindexter, et al.}{2005}]{Poindexter:2005} Poindexter S., Afonso C., Bennett D.~P., Glicenstein J.-F., Gould A., Szyma{\'n}ski M.~K., Udalski A., 2005, ApJ, 633, 914
\bibitem[\protect\citeauthoryear{Poleski \& Yee}{2019}]{Poleski:2019} Poleski R., Yee J.~C., 2019, A\&C, 26, 35
\bibitem[\protect\citeauthoryear{Rybicki, et al.}{2018}]{Rybicki:2018} Rybicki K.~A., Wyrzykowski {\L}., Klencki J., de Bruijne J., Belczy{\'n}ski K., Chru{\'s}li{\'n}ska M., 2018, MNRAS, 476, 2013
\bibitem[\protect\citeauthoryear{Shapiro \& Teukolsky}{1983}]{Shapiro:1983} Shapiro S.~L., Teukolsky S.~A., 1983, bhwd.book
\bibitem[\protect\citeauthoryear{Skowron, et al.}{2011}]{Skowron:2011} Skowron J., et al., 2011, ApJ, 738, 87
\bibitem[\protect\citeauthoryear{Smith, Mao \& Paczy{\'n}ski}{2003}]{Smith:2003} Smith M.~C., Mao S., Paczy{\'n}ski B., 2003, MNRAS, 339, 925
\bibitem[\protect\citeauthoryear{Sumi, et al.}{2011}]{Sumi:2011} Sumi T., et al., 2011, Natur, 473, 349
\bibitem[\protect\citeauthoryear{Udalski, et al.}{2008}]{Udalski:2008} Udalski A., Szymanski M.~K., Soszynski I., Poleski R., 2008, AcA, 58, 69
\bibitem[\protect\citeauthoryear{Wiktorowicz, et al.}{2019}]{Wiktorowicz:2019} Wiktorowicz G., Wyrzykowski {\L}., Chruslinska M., Klencki J., Rybicki K.~A., Belczynski K., 2019, ApJ, 885, 1
\bibitem[\protect\citeauthoryear{Wyrzykowski, et al.}{2015}]{Wyrzykowski:2015} Wyrzykowski {\L}., et al., 2015, ApJS, 216, 12
\bibitem[\protect\citeauthoryear{Wyrzykowski, et al.}{2016}]{Wyrzykowski:2016} Wyrzykowski {\L}., et al., 2016, MNRAS, 458, 3012
\bibitem[Zhu et al.(2017)]{Zhu:2017} Zhu, W., Udalski, A., Calchi Novati, S., et al.\ 2017, \aj, 154, 210 
\end{thebibliography}
\end{document}